\documentclass[nonacm, sigconf]{acmart}
\makeatletter
\def\@ACM@checkaffil{
    \if@ACM@instpresent\else
    \ClassWarningNoLine{\@classname}{No institution present for an affiliation}%
    \fi
    \if@ACM@citypresent\else
    \ClassWarningNoLine{\@classname}{No city present for an affiliation}%
    \fi
    \if@ACM@countrypresent\else
        \ClassWarningNoLine{\@classname}{No country present for an affiliation}%
    \fi
}
\makeatother

\setcopyright{none}
\renewcommand\footnotetextcopyrightpermission[1]{}

\usepackage{tikz}
\usepackage{multirow}
\usepackage{enumitem}
\usepackage{hyperref}
\usepackage[inkscapearea=page]{svg}
\usepackage{microtype}
\usepackage{caption}
\usepackage{subcaption}
\usepackage{amsmath,amsfonts}

\usepackage{titlesec}

 \AtBeginDocument{%
   \providecommand\BibTeX{{%
     \normalfont B\kern-0.5em{\scshape i\kern-0.25em b}\kern-0.8em\TeX}}}

\DeclareMathOperator*{\argmin}{arg\,min}

\author{Huwan Peng}
\email{hwpeng@uw.edu}
\affiliation{%
  \institution{University of Washington}
}

\author{Scott Davidson}
\email{stdavids@uw.edu}
\affiliation{%
  \institution{University of Washington}
}

\author{Richard Shi}
\email{cjshi@uw.edu}
\affiliation{%
  \institution{University of Washington}
}

\author{Shuaiwen Leon Song}
\email{shuaiwen.song@sydney.edu.dot.au
}
\affiliation{%
  \institution{The University of Sydney}
}
\authornote{Work done prior to affiliation with Microsoft}

\author{Michael Taylor}
\email{prof.taylor@gmail.com}
\affiliation{%
  \institution{University of Washington}
}

\begin{document}

\title[]{Chiplet Cloud: Building AI Supercomputers for Serving Large Generative Language Models}

\begin{abstract}
Large language models (LLMs) such as OpenAI's ChatGPT and Google's Gemini have demonstrated unprecedented capabilities of autoregressive AI models across multiple tasks triggering disruptive technology innovations around the world.
However, as models continue to grow the cost to serve these models also continues to grow threatening the democratization of LLMs.

To address this issue, we propose \textit{Chiplet Cloud}, a chiplet-based ASIC LLM-supercomputer architecture whose goal is to optimize the total cost of ownership (TCO) per generated token.
This architecture is a highly parameterizable ASIC and server-level architecture leveraging thousands of replicated accelerator modules collaborating to scale-up the performance of LLMs at cloud-scale.
To determine specific parameterizations of the Chiplet Cloud architecture, we implemented a two-phase hardware-software co-design methodology that can search the massive design space and fine tune the architecture across a collection of LLMs based on an accurate inference simulation.
A common bottleneck for LLMs is the memory access performance therefore we introduce CC-MEM, a scalable on-chip memory system for Chiplet Cloud architectures.
Using the CC-MEM, Chiplet Clouds can be built using only SRAMs for design points where the power and performance of memory access is critical.
The CC-MEM also includes a compression decoder module to add support for sparse models without impacting the compute units using a Store-as-Compressed, Load-as-Dense mechanism.

We evaluate Chiplet Cloud architectures across eight popular LLMs on the market representing a variety of model sizes.
Using fine tuned Chiplet Cloud servers we are able to achieve $97\times$ and $18\times$ improvement in TCO/Token over rented GPU and TPU clouds, or a $8.3\times$ and $3.7\times$ improvement over fabricated GPU and TPU clouds respectively.
Chiplet Cloud can also support $1.7\times$ larger models with a sparsity of 60\%.
\end{abstract}




\maketitle

\renewcommand{\shortauthors}{}

\section{Introduction}
Recently, generative Large Language Models (LLMs) like ChatGPT~\cite{openai_chatgpt_2022} have gained significant attention around the world due to their unprecedented ability to perform a variety of natural language tasks.
These LLMs are currently driving a technology revolution at planet-scale, changing the way we interact with AI models on a daily basis from web search, word processing, and programming~\cite{github_copilot_2023}.

\begin{figure}[t]
    \centering
    \includegraphics[width=0.49\textwidth]{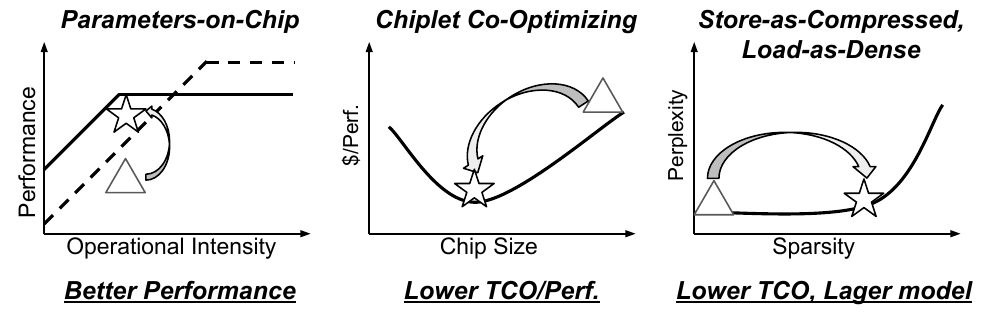}
    \caption[]{
    Compared to conventional systems, Chiplet Cloud (1) fits all model parameters inside the on-chip CC-MEM, greatly improving the performance;
    (2) co-optimizes the chip size with software mapping to reduce TCO/Perf;
    (3) exploits sparsity to reduce TCO and support larger models.
    }
    \label{fig:insights}
\end{figure}

A major contributing factor to the increase in ML capabilities comes from the unprecedented scale of the LLMs being deployed.
Most LLMs used today have billions~\cite{brown_language_2020, chowdhery_palm_2022, touvron_llama2_2023} or even trillions of parameters~\cite{fedus_switch_2021}.
Serving modern generative LLMs on commodity hardware, like GPUs, is already hitting a \textit{scalability wall}.
For example, Google Search is estimated to process over 99,000 queries~\cite{google_search_2023} per second while state-of-the-art GPT-3 throughput on GPUs is 18 tokens/sec per A100~\cite{aminabadi_deepspeed-_2022}.
If GPT-3 is embedded into every query and each query generates 500 tokens, Google would need 340,750 NVIDIA DGX servers (2,726,000 A100 GPUs) to keep up.
Assuming every GPU was able to sustain 50\% utilization, the average power would be over 1 Gigawatt which is enough energy to power 750,000 homes~\cite{cnet_gigawatt_2023}.
To address these scalability issues, we must design hardware systems that attain significantly better \textbf{\textit{total-cost-of-ownership (TCO) per token}} served.



We propose \textit{Chiplet Cloud}, a highly parameterizable chiplet-based ASIC LLM-supercomputer architecture which aims to reduce TCO per generated token.
The main insights behind the Chiplet Cloud architecture are shown in Figure \ref{fig:insights}.
To address the potential bandwidth bottlenecks of LLM inference, the Chiplet Cloud architecture allows for all model parameters and KV values to be stored in a memory system called CC-MEM (Section \ref{sec:cc_mem}), a scalable on-chip memory system for Chiplet Cloud architectures.
We use a finely tuned replicated chiplet accelerator module to reduce the fabrication cost as we scale the system to meet performance demands (Section \ref{sec:chiplet_to_cloud}).
To support models that leverage sparsity, we use a compression decoder unit which lives within the CC-MEM network to implement a \textit{Store-as-Compressed, Load-as-Dense} mechanism (Section \ref{sec:sparsity}).
We show these design choices win in the competition of TCO per token for serving generative LLMs but requires careful consideration with respect to the chiplet die size, chiplet memory capacity and bandwidth, and total number of chiplets to balance the fabrication cost and model performance (Section~\ref{sec:arch_solution} and Section~\ref{sec:design_space}).


To explore the massive hardware-software co-design space of Chiplet Cloud and find TCO per token optimal parameterizations, we propose a two-phase design-search methodology that fine tunes the architecture across a collection of LLM workloads.
The hardware exploration phase (Section \ref{sec:hardware_exploration}) conducts a bottom-up design space exploration of Chiplet Cloud hardware architecture from a flexible accelerator architecture up to a 1U rack mounted server architecture taking power budget, floorplan, and thermal constraints into account.
The software evaluation phase (Section \ref{sec:software_evaluation}) then performs a detailed performance and TCO analysis of the server designs given a specific workloads while simultaneously searching for a software mapping strategy that complements the server architecture.
While software mapping strategies for LLMs are now considered standard techniques for improving performance on existing hardware platforms, our design methodology flips the order and allows us to explore mapping strategies across all possible Chiplet Cloud hardware configurations for a software-hardware co-design methodology.

In summary, this paper makes the following contributions: 

\begin{itemize}[]
    \item We propose \textit{Chiplet Cloud}, a chiplet-based ASIC LLM supercomputer with a dedicated memory system architecture CC-MEM for serving generative LLMs designed to improve the TCO/Token over currently deployed systems (Section \ref{sec:cc_mem}, Section \ref{sec:chiplet_to_cloud});
    \item We design a compression decoder unit in the CC-MEM network and propose the store-as-compressed, load-as-dense mechanism to support sparsity enabled LLMs (Section \ref{sec:sparsity});
    \item We present a comprehensive software-hardware co-design methodology \footnote{The methodology is packaged as a design tool that can be shared to the community and is applicable to other cloud architectures.} that enables an accurate Chiplet Cloud design space exploration with software mapping optimization aware search (Section \ref{sec:methodology});
    \item We design and evaluate the Chiplet Cloud architecture across eight popular LLMs representing a variety of model sizes (Section \ref{sec:case}). Compared to running on rented GPU and TPU clouds, Chiplet Cloud can achieve up to 97$\times$ and 18$\times$ improvement in TCO/Token (Section \ref{sec:eval}).
\end{itemize}

\section{Background}
\label{sec:motivation}

\subsection{Large Generative Language Models}
\label{sec:llm_background}

\begin{figure}[t]
    \centering
    \includegraphics[width=0.48\textwidth]{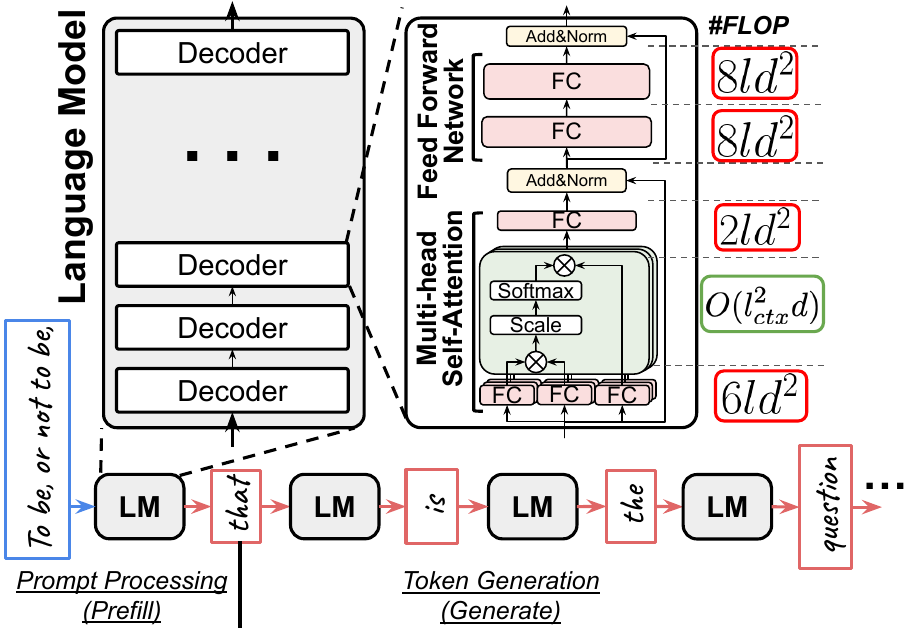}
    \caption{
            General architecture of an autoregressive generative large language model.
            In most LLMs, $d$ is significantly larger than $l_{ctx}$, causing FC layers to dominate the overall runtime.
            The inference is partitioned into two stages: prompt processing (prefill) and token generation (generate).
    }
    \label{fig:LM}
\end{figure}

While generative LLMs have undergone many improvements, the model's architecture, shown in Figure~\ref{fig:LM}, has remained relatively unchanged.
These models are constructed by stacking layers of transformer decoder blocks~\cite{vaswani_attention_2017} which are composed of a self-attention mechanism followed by a 2-layer feed-forward network. 
The number of compute operations for each step of the decoder block is shown in Figure~\ref{fig:LM}.
In modern LLMs, the fully-connected (FC) layers often dominate the runtime since the model dimension $d$ is significantly larger than the context length $l_{ctx}$, thus $O(ld^2) >> O(l_{ctx}^2d)$.
For example, in GPT-3, where $d=12288$ and $l_{ctx} \leq 4096$, more than 99\% of MAC operations are performed in the FC layers.

Generative LLM models employ autoregressive inference, as visualized in the lower part of Figure~\ref{fig:LM}.
This inference process unfolds in two stages: prompt processing, or \textit{prefill}, and token generation, or \textit{generate}.
In the prefill stage, the model ingests the entire input sequence to generate the initial output token.
In the generate stage, this output token serves as the input for subsequent passes, producing new tokens iteratively. 
This iterative process continues until a specific "end of text" token is generated or the sequence reaches a predefined maximum length. 
This research centers on optimizing token generation, which typically consumes significantly more time than the prefill phase and incurs a higher cost per token~\cite{openai_chatgpt_2022}.

Since the appearance of the original transformer architecture~\cite{vaswani_attention_2017}, the development of LLMs has mainly focused on dimension scaling, such as layer size and number of layers.
LLMs may also have different operations, such as activation function (ReLU~\cite{agarap_relu_2019}, GeLU~\cite{hendrycks_gelu_2023}, Swish~\cite{ramachandran_swish_2017} and their GLU variants~\cite{shazeer_glu_2020}), positional embeddings (absolute~\cite{vaswani_attention_2017}, relative~\cite{shaw_relative_2018}, ALiBi~\cite{press_alibi_2022} and RoPE~\cite{su_rope_2022}), and normalization (original layer normalization~\cite{ba_layernorm_2016} and RMS layer normalization~\cite{zhang_rms_2019}).

\subsection{Main Challenges}
\label{sec:challengs}

\subsubsection{\textbf{Memory Bandwidth Significantly Limits Inference Performance}}
Inference performance of a LLM, both latency and throughput, is often bottlenecked by memory bandwidth.
The FC layers are plagued with low operational intensity leading to memory bounded operation for a majority the model~\cite{aminabadi_deepspeed-_2022}. 
A prevalent strategy for enhancing throughput involves employing a large batch size, thereby improving the operational intensity of the FC layers but at the cost of an increase in prefill latency and KV cache size.
For example, a GPT-3 model with a context length of 2K would require 2 GB for the KV cache.
\emph{KV caching} is a widely adopted technique in token generation, where intermediate results of the self-attention mechanism from previous input tokens are cached, eliminating the need for recomputation. 
With a batch size of 256, the KV cache grows to 512 GB while the total model parameter size is 350 GB.
Reducing parameter and KV access latency and power is critical for achieving a good cost per throughput (i.e. TCO/Token).

\subsubsection{\textbf{Chip Costs Dominate TCO}}
The autoregressive nature of LLMs severely constrains hardware utilization.
The best hardware utilization on the state-of-the-art implementation is around 50\%~\cite{aminabadi_deepspeed-_2022} on GPUs and 40\% on TPUs (during the decoding phase)~\cite{pope_efficiently_2022}.
It is noteworthy that these are achieved with a very large batch size (e.g. 1024) as the utilization can be as low as 1\% when batch sizes are small (e.g. 4)~\cite{pope_efficiently_2022}, which is a common case for real-world LLM inference.
Compounding this issue is the massive scale of contemporary chips (e.g. A100 GPU), approaching the wafer reticle limit of around 800 mm$^2$, presenting a huge challenge in managing fabrication costs. 
Under conditions of low utilization and high chip fabrication costs, capital expenditures (CapEx) become a substantial fraction of the TCO. 
Our evaluation indicates that at 50\% utilization, the TCO of running an A100 GPU purchased at manufacturer's retail price is 97.7\% CapEx.
Even in scenarios where individuals tapeout their own GPUs, the CapEx percentage can remain as high as 58.7\%. 
The pivotal strategy for reducing TCO per generated token therefore involves addressing the challenge of capital expenditure.

\subsection{Architectural Solutions}
\label{sec:arch_solution} 
To build ASIC supercomputers for LLMs that effectively address these challenges, this paper explores an architectural solution within the following two design spaces.

\subsubsection{\textbf{New Memory System for Better Performance and TCO}}
\label{sec:sram_solution}

In situations where we are bottlenecked on memory bandwidth, our focus is to mitigate the reliance on external memory solutions such as HBM or DDR by allowing the architecture to buffer all model parameters and intermediate data (such as the KV cache) in an on-chip memory system using SRAM.
SRAM can have a much higher bandwidth and much lower access energy, albeit at the cost of lower storage density. 
Recently, there has been an industry trend to deploy more on-chip memory on deep learning accelerators to reduce the excessive off-chip memory access.
Microsoft's Brainwave~\cite{fowers_configurable_2018} pins DNN model weights in distributed on-chip SRAM.
Google's TPU v4i~\cite{jouppi_ten_2021} and TPU v4~\cite{jouppi_tpuv4_2023} contains 144 MB and 177 MB SRAM respectively, and GraphCore's IPU 2~\cite{knowles_graphcore_2021} has 896 MB SRAM.
While SRAM has better performance and access energy, it is more expensive per bit thus an exploration is required to determine if SRAM only systems are superior with respect to TCO/Token.

\subsubsection{\textbf{Chiplet for Reducing Fabrication Cost}}
\label{sec:chiplet_solution}
An extreme case of adding on-chip memory is to go wafer-scale.
Cerebaras WSE-2~\cite{systems_wafer-scale_2019} is a 46,255 mm$^2$ chip with 40 GB on-chip memory.
The niche wafer-scale designs are expensive to manufacture, resulting in limited potential for TCO reduction.
We advocate for the incorporation of chiplet technology as a pivotal strategy for managing TCO for LLM supercomputers.
Chiplet technology has recently become a new trend in the industry.
It breaks down a traditional monolithic silicon chip into multiple small chiplets and integrates them into a single package.
This approach improves fabrication yield, reduces manufacturing costs and enables die-level reuse for different system scales.
For TSMC 7nm technology with a defect density of 0.1 per cm$^2$, the unit price of a 750 mm$^2$ chip is twice that of a 150 mm$^2$ chip.
It is a currently an available commodity technology that all architects can use, aligning with our TCO/Token optimization focus.

One potential drawback of chiplet design is the high inter-chiplet communication.
Studies on GPUs and TPUs have shown that proper mapping strategies (e.g. tensor and pipeline parallelism~\cite{shoeybi_megatron-lm_2020, narayanan_efficient_2021, pope_efficiently_2022}) can effectively reduce the inter-node communication. 
Recent research~\cite{pope_efficiently_2022} proposes a \textit{2D weight-stationary layout} to partition the feed-forward network.
Compared with conventional 1D partitioning, 
this method makes the communication time of the feed-forward networks scale as $O(\frac{1}{\sqrt{n_{chips}}})$.
Systems with more chiplets benefit from this approach over systems with fewer monolithic chips.

One question in harnessing chiplet technology pertains to identifying packaging techniques that align with our design goals of reducing TCO/Token. 
Two widely adopted solutions are silicon interposers~\cite{kannan_enabling_2015} and organic substrates~\cite{naffziger_pioneering_2021}. 
Silicon interposers provide higher signal density for high bandwidth and is a key component for HBM integration.
However, it has a limited max signal distance and adds significant unit cost.
In scenarios where HBM is not employed, organic substrates could be a better choice offering a more economically viable TCO/Token.

\begin{figure*}[t]
    \centering
    \includegraphics[width=0.95\textwidth]{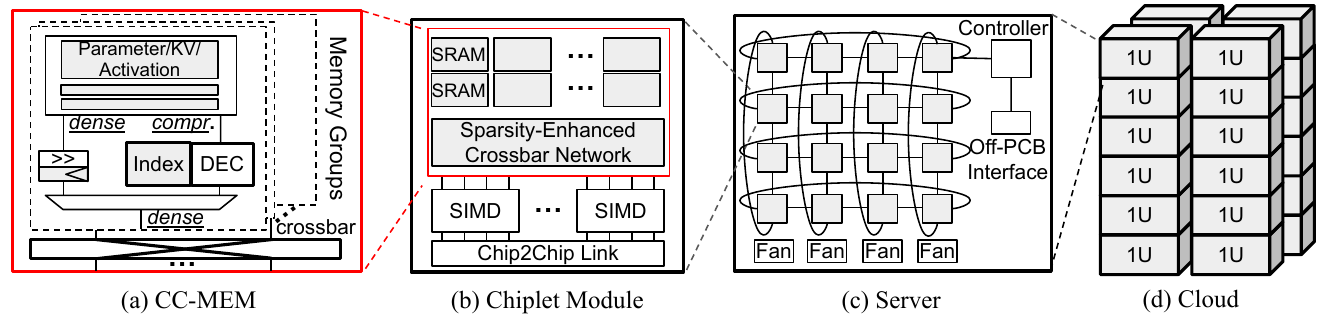}
    \caption{
    Chiplet Cloud architecture from the CC-MEM to the cloud.
    }
    \label{fig:new_arch}
\end{figure*}

\section{Chiplet Cloud: A TCO-Optimized ASIC Supercomputer Architecture for LLMs}
\label{sec:arch}

Accelerator designs often focus on raw hardware performance, however this is not always aligned with cloud hardware designers whose systems are optimized for TCO per performance \cite{jouppi_ten_2021}.
TCO includes both the the capital expenditure (\textit{CapEx}) plus the operation expenditure (\textit{OpEx}) over the lifetime expectancy of the system (\textit{Life}), giving us the equation $TCO = CapEx + Life \times OpEx$.
Optimizing TCO is therefore a balance of how much you are willing to pay for the additional performance.
Aiming for improved TCO per performance, we propose a chiplet-based cloud-scale system design for LLM inference, called \textit{\textbf{Chiplet Cloud}}.
The architectural breakdown of Chiplet Cloud is shown in Figure~\ref{fig:new_arch}, which includes the high-level abstract architecture at different levels from the memory system up to the chiplet module, server, and cloud.

\subsection{Chiplet Cloud Memory Architecture}
\label{sec:cc_mem}

The heart of Chiplet Cloud is the Chiplet Cloud Memory architecture CC-MEM (Figure~\ref{fig:new_arch} (a)).
CC-MEM is a scalable on-chip memory system with the ability to sustain high-bandwidth, low-latency read and write operations.
This is the main memory for each chiplet in the chiplet-cloud system which stores the model parameters, KV cache and activations.

The CC-MEM is designed to act as a drop-in replacement for DRAM memory but leverages SRAM to give us opportunities to take advantage of higher-bandwidth and lower-latency memory access for significantly better performance for the low operational intensity kernels of LLMs.
SRAMs are clustered into bank groups with each bank group acting as a virtual single-port memory.
Each bank group also contains a compression decode unit including a sparse tile memory. 

Bank groups are interconnected using a pipelined crossbar switching network.
The decision to use a crossbar network comes from the low-latency and low-global communication power overhead while being able to achieve a 100\% saturated throughput with reasonable network scheduling.
The biggest downside of utilizing a crossbar network comes from their area scalability as the network scales quadratically with the radices of the network.
As with many networks, this area is routing dominated.
The CC-MEM is mostly SRAM; thus, there is an abundance of routing tracks available above the SRAM devices severely lessening the area overhead of crossbar network, a concept known as \textit{NoC symbiosis}~\cite{petrisko_noc_2020}.
Crossbar networks also have the benefits of being simple to model both in terms of latency (pipeline depth) and congestion (bank conflicts).
This allows our hardware-software co-design search space to take into account memory scheduling to ensure that we can achieve the memory access performance that is required in order to hit our target TCO/performance metrics.

The CC-MEM supports burst mode operations.
Each bank group contains a simple control unit to facilitate in bursting multiple sequential read/write commands within a bank group.
This control unit is programmed using simple memory mapped control status registers.
Due to the highly structured nature of GEMM kernels, burst mode operations will make up a majority of the memory operations during moments of computation and will greatly reduce the burden on the compute unit to keep the memory system bandwidth at near-peak throughput.

\subsection{CC-MEM for Sparsity}
\label{sec:sparsity}

\begin{figure}[t]
    \centering
    \includegraphics[width=0.49\textwidth]{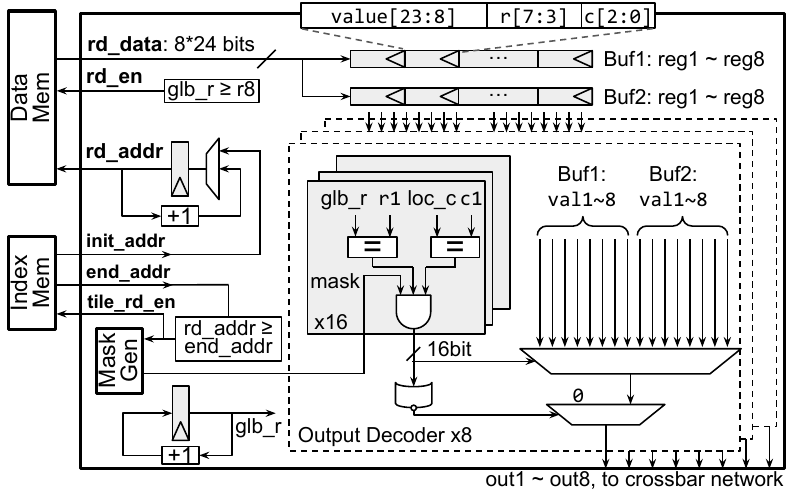}
    \caption{
    Compression decoder unit.
    }
    \label{fig:decoder}
\end{figure}

There is a growing interest in reducing LLM inference costs via model compression. 
Recent work \cite{frantar_sparsegpt_2023} has shown that large models are more compressible and have significantly less accuracy drop off than small models under compression.
OPT-175B~\cite{zhang_2022_opt}, which has the same model architecture as GPT-3, can reach 60\% unstructured sparsity with negligible increase in perplexity while requiring no fine-tuning effort.
Supporting unstructured sparse model on ASIC can be challenging since the highly irregular sparsity can lead to unpredictable data assess and compute patterns.
Simultaneously, sophisticated decoder and on-chip network architecture for sparse data dispatching can add significant area overhead.
To address these issues, we implement a \textit{Store-as-Compressed, Load-as-Dense} mechanism into the CC-MEM architecture.
Models are compressed using a tile-based compressed sparse row format~\cite{niu_tilespMv_2021} and stored in the CC-MEM in this sparse format.
However, load access patterns and data appear as if the data was stored dense.
The methodology is based on the insight that \textit{TCO/Token of our proposed system will be primarily limited by the on-chip memory size, rather than memory bandwidth and compute unit utilization}.
Reducing the required memory size will be the first priority when supporting sparse models.
Using this mechanism, the compute units are \textit{sparsity-agnostic} and do not require any special design, reducing area overhead and increasing flexibility.

To support this methodology, each bank group within the CC-MEM contains a compression decode unit.
Data in CC-MEM can be in raw dense formats or sparse compressed formats.
The decode units are controlled using a simple set of memory mapped CSRs similar to the burst mode CSRs.
Data sent over the network is always in dense formats, allowing any network attached compute units to be completely agnostic to the format the data is stored in.
Compressed data ultimately has a lower bandwidth than dense data.
This is because dense data and sparse data are both stored in the same SRAM banks which have the same peak bandwidth but sparse data has additional bits per word.

Figure~\ref{fig:decoder} shows an example design of the compression decoder unit.
In this example, the sparse matrix is divided into tiles of shape (32, 8).
Like the standard compressed sparse row format, non-zero values (NZV, 16 bits) in a tile are encoded using a 5-bit row index ($r$) and a 3-bit column index ($c$), forming a 24-bit sparse word stored in data memory.
Tile indexes are stored in a separate index memory, which is placed
together with crossbar routing tracks to minimize area overhead.
To read sparse data, the decoder sends a tile read request to the index memory and receives the initial address and end address of the NZVs in a tile.
The decoder then reads data memory at a rate of up to 8 sparse words per cycle and writes them to a double-buffer.
Depending on the row index and column index, zeros are inserted accordingly to form the original dense tile.
The unit can constantly output 8 dense words per cycle.

\subsection{From Chiplet to Cloud}
\label{sec:chiplet_to_cloud}
Figure~\ref{fig:new_arch} (b) shows a LLM accelerator chiplet module. 
Inside the chiplet, multiple SIMD cores are attached to a CC-MEM.
Compared to a fully custom compute units, the SIMD cores are more flexible with very few limitations on the types of kernels that can be efficiently supported, which is essential for supporting the various activation functions and embeddings found in modern LLMs.

In Chiplet Cloud, a single chiplet module functions as a discrete package, with multiple chiplets interlinked across the board. 
Advanced package-level solutions such the silicon interposers~\cite{kannan_enabling_2015} can provide higher signal density for high bandwidths in-package communication.
However, it has a limited reach and adds more cost.
In contrast, our Chiplet Cloud design adopts a \textit{board-level} organic substrate chiplet approach, aligning with specific communication requirements.
Given the large scale of modern LLMs, running the model within a single package of chiplets is often impractical.
Partitioning into multiple packages or even across servers becomes a necessity. 
This partitioning mandates collective operations, such as all-reduce, to occur across packages. 
Since the conventional ring all-reduce implementation is limited by the slowest link among nodes, the in-package high-speed links do not provide much help in this case.
Compared to conventional package-level chiplet, the board-level chiplet architecture eliminates cost of advanced packaging.

Each Chiplet Cloud server (Figure~\ref{fig:new_arch} (c)) contains a printed circuit board (PCB) with multiple chiplets, a controller and an off-PCB network interface. 
The controller, which can be an FPGA or a microcontroller, dispatches remote procedure calls from off-PCB interface to all chiplets.
Chiplets are connected together via a 2D torus on-PCB network, which is able to accommodate the many different mapping strategies that we might need to implement to efficiently run different models.
Candidates for chip-to-chip interfaces can be custom-designed links such as NVIDIA ground-referenced signaling GRS links~\cite{turner_ground-referenced_2018, poulton_117-pjb_2019}, Google TPU's Inter-Core Interconnect~\cite{jouppi_domain-specific_2020}, Graphcore's IPU-links~\cite{knowles_graphcore_2021}, or high-speed PCI-e which has been widely used as interconnects for many deep learning chips~\cite{smelyanskiy_zion_2019, prabhakar_sambanova_2021, liu_ai_2021}.
Off-PCB interfaces could be 10/100 Gigbit Ethernet or InfiniBand, enabling communication between adjacent servers.

\subsection{Design Space Discussion}
\label{sec:design_space}

The design space of Chiplet Cloud is a balancing act that includes many different architectural parameters across the entire system that greatly impact the resulting TCO/Token. Some aspects include
(1) \textit{Chiplet Module Size:} small chips benefit from higher yields while incurring more per-chip overhead; 
(2) \textit{Per Chiplet Memory Size:} more memory on chips means few chips required but few FLOPS per chip; 
(3) \textit{Per Chiplet FLOPS:} more FLOPS increases performance while requiring higher memory bandwidth, resulting in a larger memory crossbar; 
and (4) \textit{Software Mapping:} the trade-off between different parallelisms affects utilization and interconnect data communication.
Since all of these aspects are tightly coupled, a comprehensive design methodology is critical to optimize the end-to-end performance and TCO.


\section{Chiplet Cloud Design Methodology}
\label{sec:methodology}

\begin{figure}[t]
    \centering
    \includegraphics[width=0.50\textwidth]{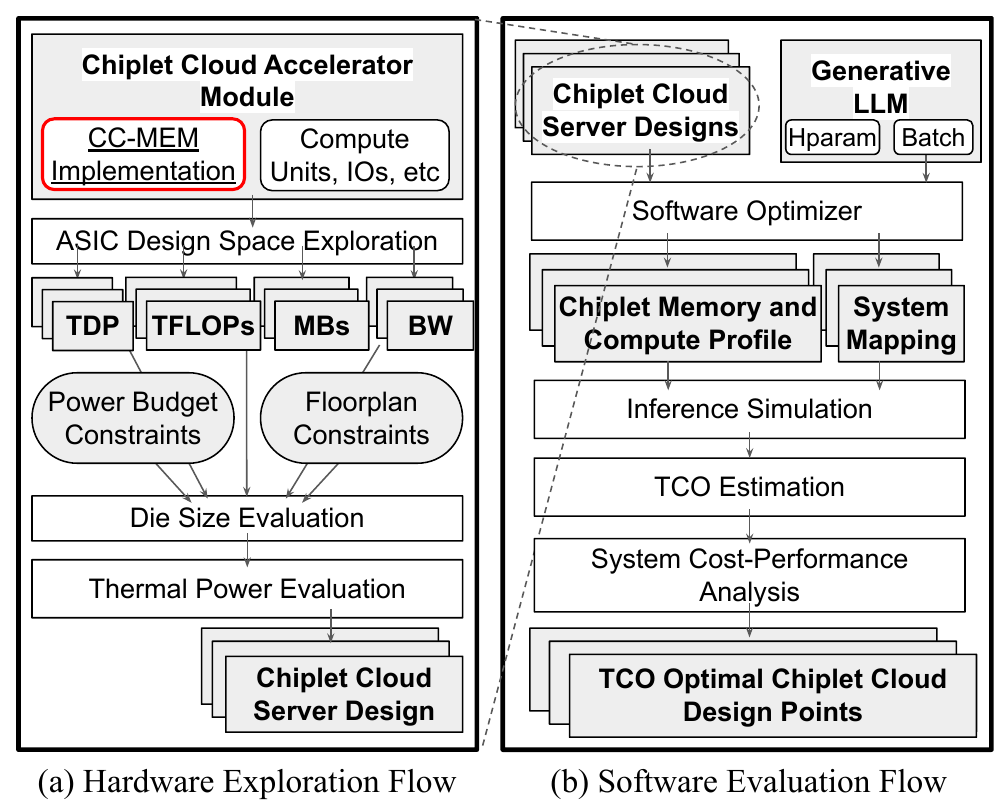}
    \caption{
        Two phase design methodology flow diagram.
        (a) The hardware exploration flow performs a bottom-up, LLM agnostic design space exploration generating thousands of realizable Chiplet Cloud server designs.
        (b) The software evaluation flow then takes the realizable server design points along with a generative LLM specification to perform software optimized inference simulations and TCO estimations to find the optimal Chiplet Cloud design points.
    }
    \label{fig:design_methodology_flow}
\end{figure}

The design methodology for Chiplet Cloud, shown in Figure~\ref{fig:design_methodology_flow}, is a critical component for finding design points with best-in-class TCO/Token.
When we scale to the size of cloud supercomputers, small changes at the architecture level have massive implications at the warehouse scale over the lifetime deployment of the system.
We there take a brute-force approach to the design space exploration which eliminates any preconceived assumptions and allows the models to decide what is the best hardware-software design pair.

\subsection{Phase 1: Hardware Exploration}
\label{sec:hardware_exploration}

The hardware exploration phase of the Chiplet Cloud design methodology (as shown in Figure~\ref{fig:design_methodology_flow}(a)) is a bottom-up, LLM agnostic, design space exploration resulting in \textit{tens of thousands} of feasible Chiplet Cloud server designs.
This exploration manifests as a substantial parameterization sweep, guided by multi-level hardware specifications provided as input, and user-preconfigured constraints and constants. 
Some essential parameters are listed in Table~\ref{tab:parameters}.
Throughout the hardware exploration, every conceivable combination of input values undergoes evaluation, spanning diverse parameters like chip performance, SRAM size, and variations in the number of chiplets within servers.

%
%

\textbf{\underline{ASIC Design Space Exploration.}}
Since our optimization target is TCO/Token, the 3 most critical estimations that need to be performed are the silicon area, power consumption, and end-to-end performance, as these are the 3 biggest contributors to capital-expenditure, operational-expenditure, and tokens-per-second respectively.
The silicon area, energy per operation, memory bandwidth, and peak power draw of the design are estimated during the hardware exploration phase, and the end-to-end performance is evaluated in the software evaluation phase with a specific application.

\textbf{\underline{Die Size Evaluation.}}
The die area modeling is separated into memory, compute, and auxiliary components.
For memory, we use the CC-MEM architecture described in Section~\ref{sec:cc_mem}, including multiple SRAM banks, sparse decoders, and a crossbar network.
It was modeled using a 12nm implementation, which is synthesized, placed and routed using Synopsys Design Compiler and Synopsys IC Compiler II.
These results were then scaled to 7nm using 2 scaling factors, one for area attributed to SRAM bitcells which uses the reported High-Density SRAM bitcell area values for 7nm and one scaling factor for routing dominated area which uses reported contacted-poly-pitch by minimum-metal-pitch (CCP-MMP) \cite{wikichip_7nm_2022}.
The CC-MEM is routing dominated in regions that are not SRAM therefore we do not need to take into consideration transistor scaling.
The computation and auxiliary component area modeling and constraints are derived from publicly available information on 7nm NVIDIA A100 GPU \cite{nvidia_a100_2023}.
Compared to a fully custom accelerator designs that are prolific in the literature, the A100 is a more flexible offering with very few limitations in the types of kernels it can operate efficiently on, which is essential for supporting various activation functions and positional embeddings, as introduced in Section~\ref{sec:llm_background}. 

\begin{table}[t]
\caption{Essential parameters of Chiplet Cloud hardware exploration}
\centering
\begin{tabular}{|l|l|}
\hline
\textbf{Parameters}       & \textbf{Range}                    \\ \hline \hline
Technology                & 7nm                               \\ \hline
Die Size                  & 20 mm$^2$ to 800 mm$^2$           \\ \hline
Memory Density            & Scaled from 12nm implementation   \\ \hline
Memory Bandwidth          & Scaled from 12nm implementation   \\ \hline
Compute Density           & 2.65 mm$^2$/TFLOPS                \\ \hline
Power Density             & 1.3 W/TFLOPS, \textless 1W/mm$^2$ \\ \hline
Chip IO                   & 25 GB/s * 4 links                 \\ \hline 
Wafer Defect Density      & 0.1 / cm$^2$                      \\ \hline
Wafer Cost                & \$10000                           \\ \hline
\hline
Server Size               & 1u, 19 inch                       \\ \hline
Lanes per Server          & 8                                 \\ \hline
Silicon per Lane          & \textless 6000 mm$^2$             \\ \hline
Chips per Lane            & 1 to 20                           \\ \hline
Power per Lane            & \textless 250 W                   \\ \hline
Server Thermal            & Adapted from ASIC Clouds \cite{magaki_asic_2016}   \\ \hline
PSU Efficiency            & 0.95                              \\ \hline
DCDC Efficiency            & 0.95                              \\ \hline
Ethernet                  & 100 Gigabit, \$450                \\ \hline
Server Life               & 1.5 year                          \\ \hline

\end{tabular}
\label{tab:parameters}
\end{table}

\textbf{\underline{Thermal Power Evaluation.}} 
The power model is also derived from A100 GPU.
We normalize the TDP and peak performance to W/FLOPS and use that value to model our Chiplet Cloud accelerator.
This is a simple and conservative estimate since a significant portion of GPU power comes from DRAM, but it greatly simplifies the power model and increases the confidence interval of our TCO/Token modeling results.
Combined with the area model, we limit the chip power density to be no more than 1 W/mm$^2$.
On the server level, we will further refine the peak power density limitations based on the full-server thermal analysis, and eliminate any thermally infeasible designs.


\subsection{Phase 2: Software Evaluation}
\label{sec:software_evaluation}

The second phase of the design methodology models the execution time of specific workloads across the hardware design points and searches for optimal Chiplet Cloud architectural configurations, as shown in Figure~\ref{fig:design_methodology_flow}(b).

\textbf{\underline{Software Optimizer.}} 
In this stage, we conduct software optimizations, incorporating \textit{tensor parallelism} and \textit{pipeline parallelism} with microbatch tuning~\cite{shoeybi_megatron-lm_2020, narayanan_efficient_2021}. 
It will first look at the hyper-parameters of the LLM, such as the model dimension $d_{model}$, number of layers, context length, attention mechanism type (multi-head or multi-query), as well as expected batch size.
Then it will decompose the full model into a collection of small kernels that can be mapped to the individual chiplets throughout the system.
In cases where the model cannot fit into a single server, the server will be replicated to scale up the entire system until there are enough resources to execute the application.
This results in a system mapping which has the portion of the model that each chiplet in the whole system will be responsible for executing.
There also exists a chiplet memory profile (required memory for weights, activations, and the KV cache) and chiplet compute profile (operation type and size) for the portion of the model that will be running on the individual chiplet, which will allow us to accurately model the end-to-end performance of the full system.

\textbf{\underline{Inference Simulation.}} 
The comprehensive end-to-end inference simulation for the Chiplet Cloud system initiates with an analytical analysis of the compute kernel, derived from the compute profile, and memory access kernel, derived from the memory profile.
The size and configuration of these operations are scrutinized at the microarchitectural level to determine the corresponding latencies and energy.
Given the system's distributed nature across multiple chiplets, it becomes imperative to model data communication overhead, including all-reduce operations latency and energy.
The latency of an all-reduce operation involving $B$ bytes of data across $N$ nodes is disassembled into one reduce-scatter and one all-gather operation, both sharing the same latency
$$T_{reduce-scatter}=(N-1)\frac{D/N}{B}+T_{init}$$
where $B$ denotes the bandwidth of the slowest connection among the nodes, and $T_{init}$ represents the time required to initialize the operation.

\begin{figure}[]
    \centering
    \includegraphics[width=0.49\textwidth]{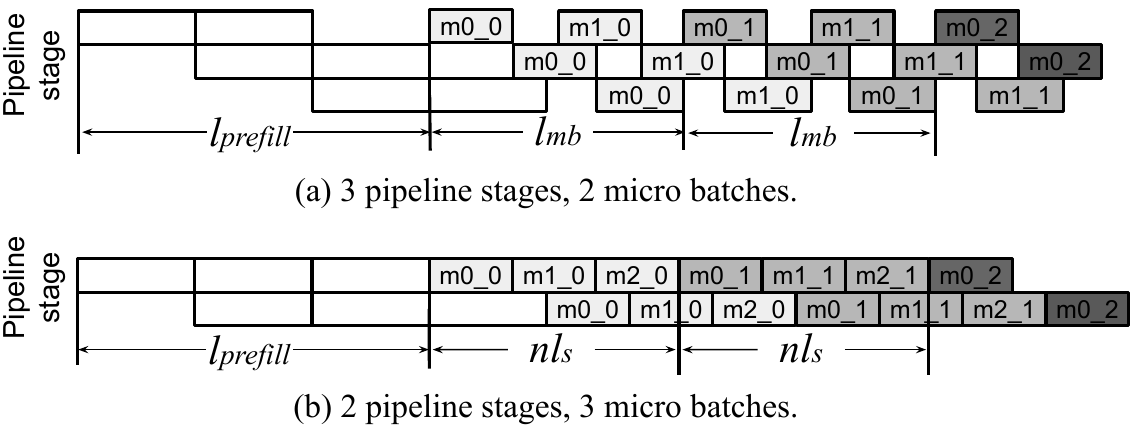}
    \caption{
    Diverse scheduling strategies for pipeline-parallelism and micro-batching.
    White boxes represent prompt processing (prefill), while shaded boxes denote token generation. The token generation throughput can be constrained either by (a) the micro-batch latency $l_{mb}$ or (b) the product of the micro-batch count and the pipeline stage latency $nl_{s}$.
    }
    \label{fig:pipeline}
\end{figure}

Summing the latencies of all kernels and communications yields the latency of a single micro-batch inference. 
The subsequent step involves determining the end-to-end latency and throughput.
Aligning with the pipeline schedule and micro-batching strategy akin to DeepSpeed Inference~\cite{aminabadi_deepspeed-_2022}, we aim to enhance system utilization and mitigate pipeline bubbles, as depicted in Figure~\ref{fig:pipeline}.
Assuming a micro-batch latency denoted as $l_{mb}$ and a pipeline stage latency of $l_{s}$, with a total of $n$ micro-batches in consideration, the per-token latency during generation is constrained by both $l_{mb}$ and the product of the micro-batch count and the pipeline stage latency, i.e., $nl_{s}$. 
The overall latency for generating $t$ tokens is given by the formula:
$$l_{all}=l_{prefill}+(t-1)\max(l_{mb}, nl_{s})$$
The throughput for a batch size $N$ is expressed as
\begin{align*}
throughput & = Nt/l_{all} \\ 
           & = \frac{Nt}{l_{prefill}+(t-1)\max(l_{mb}, nl_{s})} \\
           & \approx \frac{N}{\max(l_{mb}, nl_{s})}
\end{align*}
Given that the total generation time is often significantly longer than the prefill time, and the token count $t$ is typically substantial, we can neglect these factors. 
Therefore, the thoughtful selection of pipeline parallelism and micro-batching scheduling strategies is crucial for sustaining high system utilization and concurrently improving both latency and throughput.

In the context of systems like Chiplet Cloud, assuming a unit generation latency for batch size $N=1$ and pipeline $p=1$ as $\tau$, and recognizing that the system is often compute-bound, the micro-batch latency is expressed as $l_{mb}=\frac{N}{n}\tau$, and the single stage latency becomes $l_{s}=\frac{l_{mb}}{p}=\frac{N}{np}\tau$.
For a given system and batch size $N$, our objective is to determine the pipeline stages $p$ and micro-batch counts $n$ that maximize throughput. This is achieved by minimizing the expression
\begin{align*}
\argmin_{n,p} \max(l_{mb,g}, nl_{s,g}) &=
\argmin_{n,p} \max(\frac{N}{n}\tau, \frac{N}{p}\tau)\\
&=\argmin_{n,p} \max(\frac{1}{n}, \frac{1}{p})
\end{align*}
Subject to the constraints $n\leq N$ and $p\leq\#layers$.
Consequently, our goal is to maximize both $p$ and $n$ to optimize throughput.
Hence, it is limited by the minimum of the number of layers and the batch size $N$. 
While it is acknowledged that practical considerations, such as communication latency and hardware inefficiencies, may introduce deviations from this idealized scenario, our approach provides a valuable starting point for exploring optimal mapping strategies.


\textbf{\underline{TCO Estimation.}}
The TCO model is based the model by Barroso et al~\cite{barroso_datacenter_2013}, which includes both capital expenditures (\emph{CapEx}) and operating expenses (\emph{OpEx}) from the system as well as the datacenter hosting the system. 
The \emph{CapEx} includes the silicon die cost, package cost, PCB cost, power supply unit cost,  heatsink cost, fan costs, Ethernet controller cost, and control processor cost.
The \emph{OpEx} is calculated based on the power consumption of the system.
Based on the system TDP and utilization from inference simulation, the full system average power consumption is used to determine the power draw from the silicon dies of the Chiplet Cloud accelerators.
To estimate the die cost, we first calculate the number of fully patterned dies per wafer (DPW).
This is the number of rectangular dies with the given die size dimensions that we can slice out of a traditional 300mm circular wafer.
Cost per die is then calculated as
$$cost_{die}=(\frac{cost_{wafer}}{DPW}+cost_{test})/Y_{die}$$
Where ${cost_{wafer}}$ is wafer price, $cost_{test}$ is testing cost, and $Y_{die}$ is die yield.
We use the classical negative binomial model~\cite{cunningham_yield_1990} for yield which is as follow
\begin{equation*}
    Y_{die}=(1+\frac{AD_0}{\alpha})^{-\alpha}
\end{equation*}
Where $A$ is die area, $D_0$ is defect density and $\alpha$ is cluster parameter.
Since manufacturing yields drop with chip area, it makes economic sense to design smaller chips.


\textbf{\underline{System Cost-Performance Analysis.}} 
The end-to-end performance and the corresponding TCO are fed into the system cost-performance analysis engine, where we compute all TCO related metrics and output the optimal design points under different hardware and software constraints.

\subsection{Generalizing the Design Methodology}

While this work is focused on trying to find cloud-scale architectures with best-in-class TCO/Token performance, the methodology of designing scale-up cloud systems is still applicable to existing ASIC architectures or architectures designed for programmable devices such as CGRAs or FPGAs.
Given an appropriate power, performance and area estimation model for the accelerator module, this methodology is applicable.

\section{Case Studies}
\label{sec:case}

\begin{table*}[t]
\caption{TCO/Token optimal Chiplet Cloud systems for different language models. }
\centering
\resizebox{0.99\textwidth}{!}{%
\begin{tabular}{|l|c|c|c|c|c|c|c|c|}
\hline
Model                  & \textbf{GPT-2~\cite{radford_language_2019}} & \textbf{Megatron~\cite{shoeybi_megatron-lm_2020}} & \textbf{GPT-3~\cite{brown_language_2020}} & \textbf{Gopher~\cite{rae_gopher_2022}} & \textbf{MT-NLG~\cite{smith_using_2022}} & \textbf{BLOOM~\cite{bigscience_bloom_2023}} & \textbf{PaLM~\cite{chowdhery_palm_2022}} & \textbf{Llama-2~\cite{touvron_llama2_2023}} \\ \hline \hline
Parameters (B)         & 1.5    & 8.3     & 175     & 280    & 530    & 176    & 540    & 70      \\ \hline
d$_{model}$            & 1,600  & 3,072   & 12,288  & 16,384 & 20,480 & 14,336 & 18,432 & 8,192   \\ \hline
Layers                 & 48     & 72      & 96      & 80     & 105    & 70     & 118    & 80      \\ \hline
\hline
Die Size (mm$^2$)      & 60     & 40      & 140     & 100    & 160    & 120    & 100    & 80      \\ \hline
MB per Chip            & 32.8   & 27.0    & 225.8   & 151.0  & 198.0  & 137.5  & 95.0   & 82.5    \\ \hline
TFLOPS per Chip        & 5.60   & 2.87    & 5.50    & 4.83   & 6.32   & 7.02   & 12.07  & 7.62    \\ \hline
BW per Chip (TB/s)     & 2.80   & 2.29    & 2.75    & 2.41   & 4.21   & 3.51   & 1.51   & 1.90    \\ \hline \hline
Chips per Server       & 128    & 144     & 136     & 160    & 160    & 152    & 120    & 72      \\ \hline
Number of Servers      & 24     & 8       & 96      & 80     & 105    & 70     & 118    & 80      \\ \hline \hline
Tensor Parall. Size    & 64     & 144     & 136     & 160    & 160    & 152    & 120    & 72      \\ \hline
Pipeline Parall. Size  & 48     & 8       & 96      & 80     & 105    & 70     & 118    & 80      \\ \hline
Batch Size             & 128    & 8       & 256     & 128    & 128    & 128    & 1024   & 512     \\ \hline
Micro-Batch Size       & 2      & 1       & 2       & 2      & 1      & 2      & 8      & 4       \\ \hline \hline
Max Context Length & 16K  & 256K   & 8K   &  16K    & 16K    & 16K    & 2K    & 4K    \\ \hline
Tokens/Sec per Chip    & 473.3  & 69.7    & 8.1     & 4.3    & 2.7    & 8.6    & 7.0    & 26.5    \\ \hline
TCO/1M Tokens (\$)     & 0.001 & 0.008  & 0.161  & 0.228 & 0.521 & 0.141 & 0.245 & 0.046  \\ \hline
\end{tabular}
}
\label{tab:all_models}
\end{table*}

To evaluate Chiplet Cloud and the design methodology, we performed a case study on eight language models, including GPT-2~\cite{radford_language_2019}, Megatron-LM~\cite{shoeybi_megatron-lm_2020}, GPT-3~\cite{brown_language_2020}, Gopher~\cite{rae_gopher_2022}, MT-NLG~\cite{smith_using_2022}, BLOOM~\cite{bigscience_bloom_2023}, PaLM~\cite{chowdhery_palm_2022} and Llama-2~\cite{touvron_llama2_2023}.
Details about these models are shown in Table \ref{tab:all_models}.
All studies were conducted on publicly released data, such as model architecture hyper-parameters, and do not use actual weights.

\subsection{Design Space Exploration}

We performed a thorough design exploration 
under 3 different context length scenarios (1024, 2048 and 4096) and on batch sizes from 1 to 1024.
This exploration results in over 2 million valid design points for each model.
Each design point combines the result from both hardware exploration and software evaluation, which includes hardware design (chip and server), software mapping (tensor parallelism size, pipeline parallelism size, batch size and micro-batch size), cost (OpEx and CapEx) and performance (latency and throughput), etc.

Table~\ref{tab:all_models} shows the TCO/Token optimal Chiplet Cloud designs for each model in our case study.
We found that all TCO-optimal designs are targeting batch sizes greater than or equal to 32.
Large batch sizes are good for utilization in FC layers but will require additional silicon for memory to account for a larger KV cache.
This means we either need bigger chips which greatly increase CapEx, or more chips which generate more inter-node traffic and hurt throughput.
This will either results in larger chips which will sharply increase the CapEx as our silicon per chip gets larger and yield gets worse, or it will generate systems with a larger number of chips increasing the amount of inter-chip communication and diminishing the end-to-end performance.
Finding batch sizes that balance each factor is essential to achieve good TCO/Token but is challenging to find.
Each optimal design points across our 8 models all have different chip, server designs, and mapping strategies demonstrating the importance of our design methodology---every aspect of the system affects performance and cost and are sensitive to the requirements of the workload.

While the workload does impact the optimal Chiplet Cloud configuration, this doesn't mean that a Chiplet Cloud instance can only run a single model.
Additional discussion on the impact of running non-optimized models and how a multi-model objective optimization perform can be found in Section \ref{sec:flexibility}.

\subsection{Design Insights}

\begin{figure}[t]
    \centering
    \includegraphics[width=0.5\textwidth]{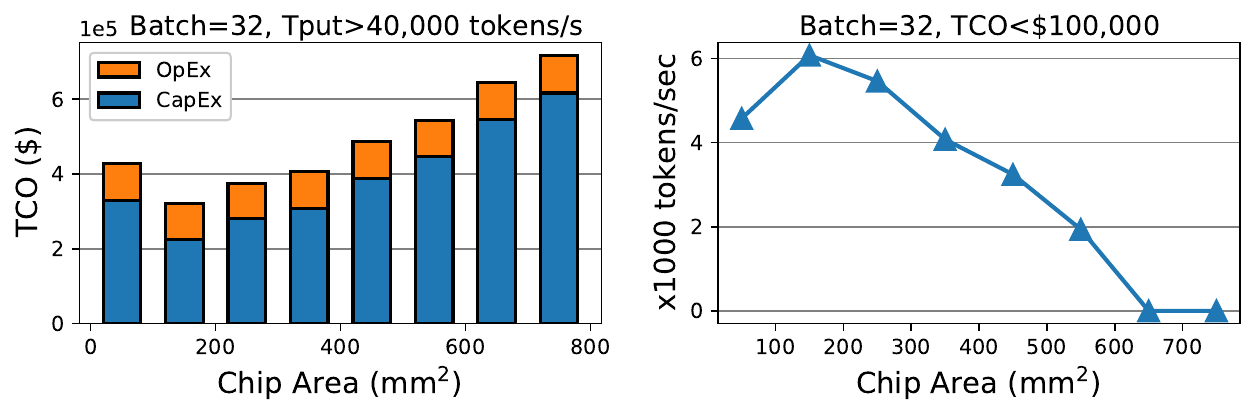}
    \caption{
    Proper chip size can reduce the fabrication costs (CapEx) without compromising performance as much. 
    Left: For a given throughput requirement, chips with a size of less then 200 mm$^2$ have lowest TCO.
    Right: For a given TCO budget, chips with a size between 100 mm$^2$ to 200 mm$^2$ achieve the best throughput.
    }
    \label{fig:chip_exploration}
\end{figure}

\textbf{How chip size affects TCO and performance.}
Figure ~\ref{fig:chip_exploration} shows the results of GPT-3 in two different scenarios.
On the left is how we should choose the die size to lower TCO for a given minimum throughput requirement.
Compared to chips over 700 mm$^2$, which is the size of many traditional large monolithic chips, a chip around 200 mm$^2$ reduces TCO by about 2.2$\times$ and still meets the throughput constraint.
We also find the CapEx exceeds 80\% of TCO for most designs.
The right side of Figure~\ref{fig:chip_exploration} shows chips with a size between 200 mm2 to 300 mm2 achieve the best throughput for a given TCO budget.
This shows that proper chip sizing can effectively reduce TCO without compromising performance.

\begin{figure}[t]
    \centering
    \includegraphics[width=0.505\textwidth]{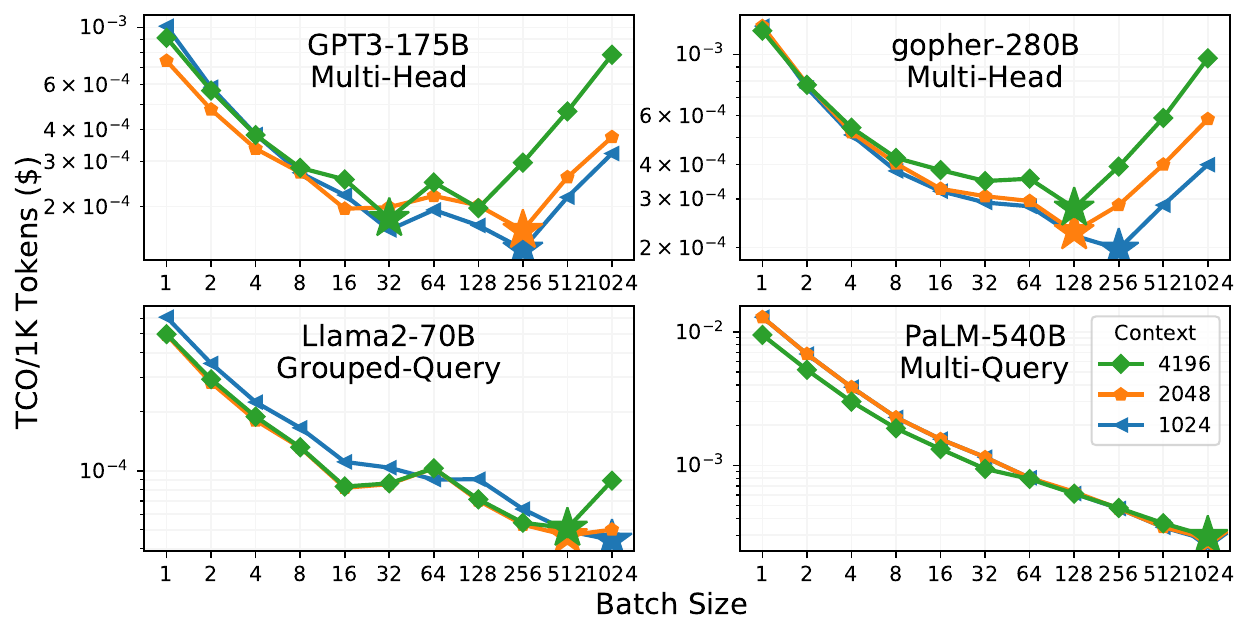}
    \caption{
    The optimal TCO/Token under different batch sizes.
    Small batch requires less silicon, and large batch benefits weight reuse.
    The optimal batch size for multi-head models is between 32 to 256, while the multi-query and grouped-query models are able to maintain a near-optimal TCO/Token at batch size 1024.
    }
    \label{fig:batch_size_all_models}
\end{figure}

\textbf{How the batch size affects TCO/Token.}
Figure~\ref{fig:batch_size_all_models} shows the TCO/1K Tokens versus batch size across 4 models and 3 context lengths.
When the batch size is increased from 1, 
TCO/Tokens improves due to increases in compute utilization by providing
more opportunities to exploit pipeline parallelism.
As the batch size continues to increase, the utilization will reach a peak.
For the traditional multi-head model, more silicon is required for KV cache in large batch size and long contexts, which significantly increases TCO/Token.
Chiplet Cloud supports batch sizes up to 128 with near-optimal TCO/Token for these models.
PaLM adopts multi-query attention~\cite{shazeer_mqa_2019} and Llama-2 adopts grouped-query attention~\cite{ainslie_gqa_2023}, where key and value are shared across all or some groups of attention heads, which reduces the size of the KV cache by a factor of number of heads. 
For these models, Chiplet Cloud supports batch sizes up to 1024 with near-optimal TCO/Token.
The cost of longer contexts is negligible, especially when the batch size is not too large.

\begin{figure}
    \centering
    \includegraphics[width=0.49\textwidth]{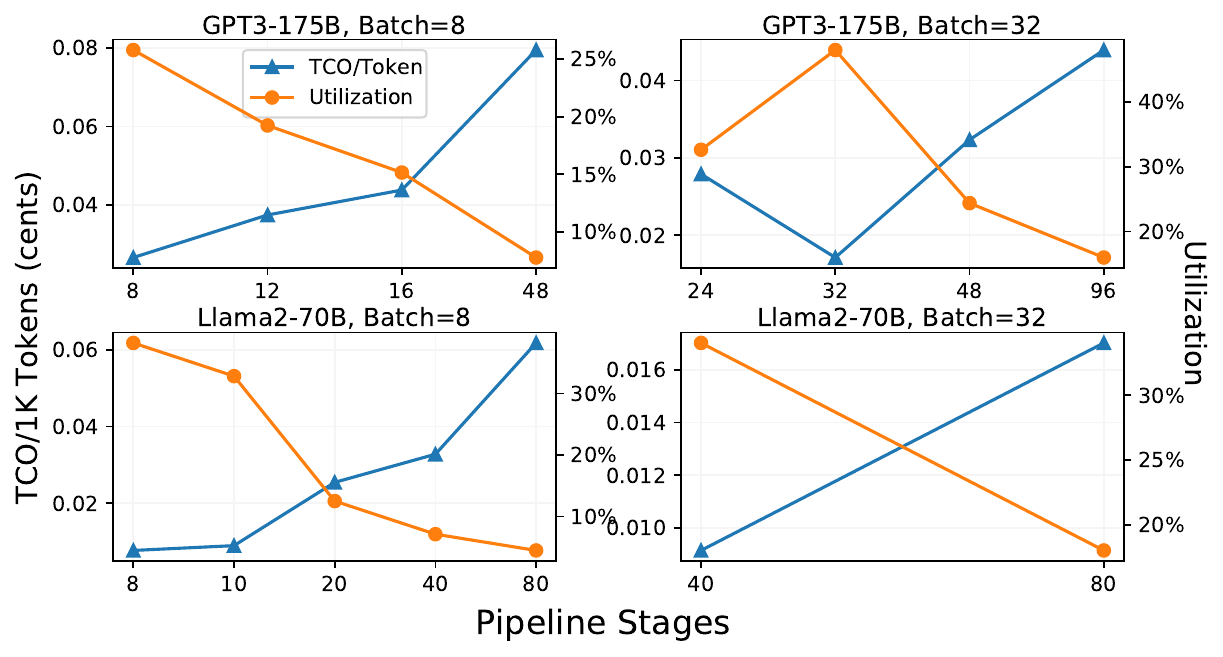}
    \caption{
    Pipeline stages sweeping for different models and batch sizes.
    The number of pipeline stages close to the batch size usually achieves the highest utilization, resulting in the optimal TCO/Token.
    }
    \label{fig:p_sweep}
\end{figure}

\textbf{How the mapping strategy affects TCO/Token for a given batch size.}
Figure~\ref{fig:p_sweep} shows that when the number of pipeline stages $p$ (i.e. the pipeline parallelism size) is close to the batch size, the system utilization is the largest and TCO/Token is optimal.
When these two numbers are similar, the system can take full advantage of pipeline parallelism with a micro-batch size of 1, so the number of micro-batches is also close (if not equal) to the pipeline stage~\cite{aminabadi_deepspeed-_2022}.
This helps balance the latency of micro-batches passing through all pipeline stages and pipeline stages completing all micro-batches.
This finding aligns with our inference simulation discussed in Section \ref{sec:software_evaluation}. Specifically, when $p$ is too small, performance is hindered by pipeline stage latency, whereas an excessively large $p$ results in limitation by microbatch latency. 

\section{Evaluation}
\label{sec:eval}
In this section, we evaluate the performance and cost of Chiplet Cloud for serving large language models.
The key metric we are targeting is \textit{TCO/Token}.
TCO/Token is measured as cost per token generated and is the key factor in the ability to democratize LLMs.
One of the most popular business models for generative LLMs is also to charge users per generated token.
Lower TCO/Token not only adds more profit margins, but also makes LLMs more approachable.
We compare Chiplet Cloud to state-of-the-art GPU and TPU cloud implementations.
We also evaluate the sparsity support and flexibility of Chiplet Cloud architectures.



\subsection{Comparison with GPUs and TPUs. }

\begin{figure}[t]
    \centering
    \includegraphics[width=0.43\textwidth]{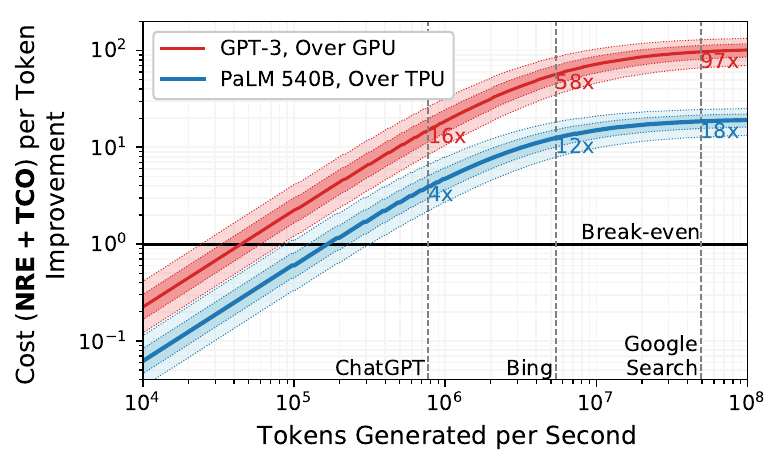}
    \caption{
    Compared to A100 GPU and TPUv4, Chiplet Cloud can achieve over $97\times$ and $18\times$ improvement in (NRE+TCO)/Token on GPT-3 and PaLM 540B, respectively. 
    The light and dark shaded regions represent the results 
under $\pm30\%$ and $\pm15\%$ input variance.
    }
    \label{fig:cc_tpu_gpu}
\end{figure}

We compare optimal Chiplet Cloud designs from Section \ref{sec:case} to state-of-the-art A100 GPU~\cite{aminabadi_deepspeed-_2022} and TPUv4~\cite{pope_efficiently_2022} implementations.
Neither work is specifically optimized for TCO/Token. For our comparison, we choose the throughput optimal result for GPU, and the utilization optimal result for TPU, which are key indicators that you are close to TCO/Token optimal.
Compared to GPU and TPU clouds, our design achieves up to $106.0\times$ and $19.9\times$ TCO/Token improvement on GPT-3 and PaLM 540B respectively.
TCO for GPUs and TPUs are based on the best cloud rental price we could find~\cite{tpu_price_2023,lambda_price_2023}.

Adding the NRE of Chiplet Cloud (\$35M, estimated based on the NRE model from Moonwalk~\cite{khazraee_moonwalk_2017}), we show the actual cost improvement in Figure~\ref{fig:cc_tpu_gpu}. 
As the number of tokens expected to be generated (x-axis) grow, NRE is greatly amortized and Chiplet Cloud gains more improvement over GPU and TPU.
Compared to A100 GPU and TPUv4 clouds, at the scale of Google search (99,000 queries per second~\cite{google_search_2023}, and assuming 500 tokens per query), Chiplet Cloud achieves $97 \times$ and $18\times$ improvement on (TCO+NRE)/Token, respectively.
We also add variance to 2 inputs that are difficult to accurately estimate, those being the TCO of GPU and TPU clouds, and the NRE of Chiplet Cloud.
With a $\pm30\%$ variance of these inputs, Chiplet Cloud is still expected to maintain a $66\times$ to $129\times$ improvement over GPU, and $12\times$ to $24\times$ improvement over TPU.



\begin{figure}[t]
    \centering
    \includegraphics[width=0.50\textwidth]{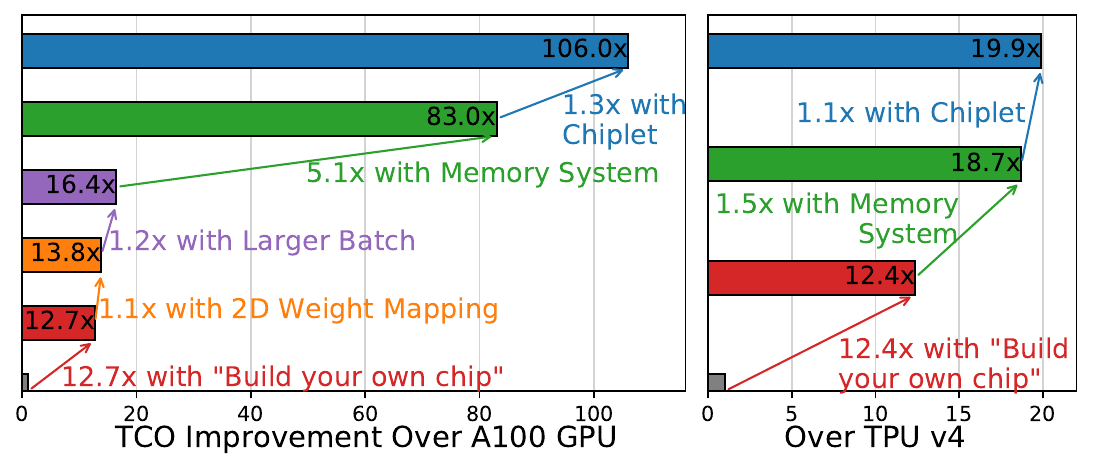}
    \caption{
    TCO/Token improvement breakdown over GPU and TPU. 
    }
    \label{fig:perf_breakdown}
\end{figure}

Figure~\ref{fig:perf_breakdown} shows the breakdown of TCO/Token  of Chiplet Cloud over GPU and TPU.
Some of the improvement in TCO comes from building the silicon instead of renting it.
To analyze the impact of owning a chip, we feed the chip and server specifications of A100 and TPU v4 into our TCO model.
The results show that owing the chip saves $12.7\times$ and $12.4\times$ in TCO/Token.
Note that the actual savings should be less than this, as our model does not include the cost of liquid cooling and advanced packaging, which are critical for TPUs and GPUs but not required for Chiplet Cloud.
We see that our specialized memory system improves TCO/Token by $5.1\times$ and $1.5\times$ over GPUs and TPUs, while die sizing improves it by an additional $1.3\times$ and $1.1\times$.
Compared to GPUs, the 2D weight-stationary layout in feed-forward network and the larger batch sizes lead to a $1.1\times$ and $1.2\times$ improvement respectively.
Both of these optimizations are supported in the TPU implementation.

\begin{figure}[t]
    \centering
    \includegraphics[width=0.45\textwidth]{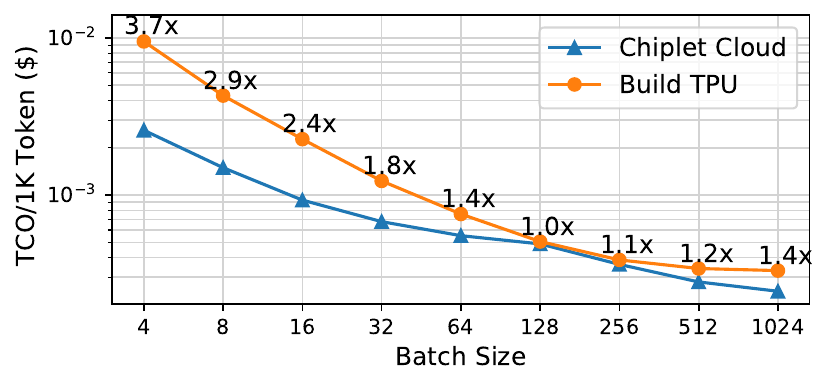}
    \caption{
    Chiplet Cloud is more efficient than TPU v4 at most batch sizes, especially for small batch sizes.
    TPU performance is from~\cite{pope_efficiently_2022} with and TCO is from our model.
    }
    \label{fig:cc_tpu_batch}
\end{figure}

In Figure~\ref{fig:cc_tpu_batch}, we compare the architectural benefits of Chiplet Cloud versus TPU v4~\cite{pope_efficiently_2022}
using our model for the TPU's TCO.
Chiplet Cloud is more efficient at most batch sizes and achieves a TCO/Token improvement of up to $3.7\times$ at batch size 4 as the high-bandwidth CC-MEM benefits from low operational intensity.

\subsection{Sparse Models Evaluation}

\begin{figure}
    \centering
    \includegraphics[width=0.45\textwidth]{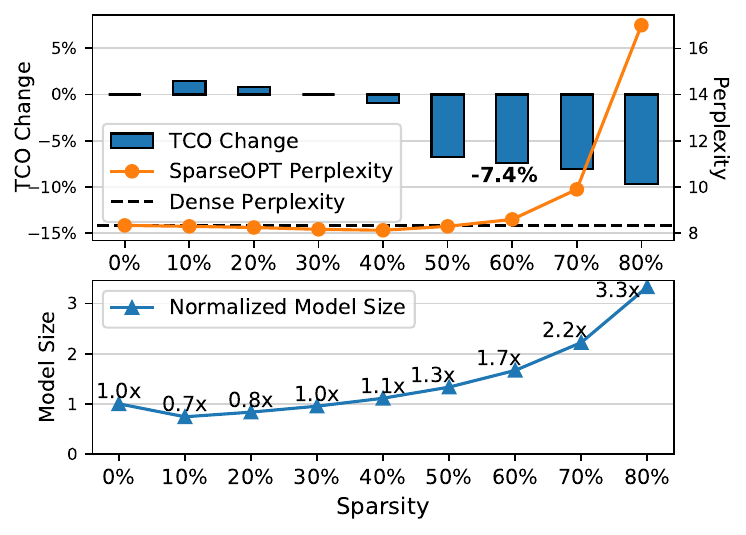}
    \caption{
    Top: TCO/Token and perplexity (from SparseGPT \cite{frantar_sparsegpt_2023}, lower is better) of OPT-175B under different sparsity. 
    Chiplet Cloud can further reduce 7.4\% of TCO/Token at 60\% sparsity with negligible increase in perplexity.
    Bottom: Chiplet Cloud supports a $1.7\times$ larger model with a sparsity of 60\%.
    }
    \label{fig:sparse_models}
\end{figure}

We evaluate the sparse models in Figure~\ref{fig:sparse_models}.
The top plot compares TCO and perplexity of OPT-175B~\cite{zhang_2022_opt} under different weight sparsities.
The perplexity values are from SparseGPT~\cite{frantar_sparsegpt_2023}.
The blue bars show the change in TCO/Token compared to using the non-compressed dense model. 
At low sparsity (such as 10\% and 20\%), TCO/Token increases because it requires more memory to store compressed format encoding overhead. 
60\% sparsity represents a sweet spot where the perplexity of the model is only marginally above that of the dense model while attaining a 7.4\% improvement in TCO/Token.
Additional sparsity continues to give additional improvements in TCO but the model perplexity starts to increase rapidly.
Chiplet Cloud also supports larger models with sparsity.
The bottom of Figure~\ref{fig:sparse_models} shows that under the same system configuration, Chiplet Cloud is able to support models with $1.7\times$ parameters at a sparsity of 60\%. 

\subsection{Chiplet Cloud Flexibility}
\label{sec:flexibility}

Flexibility is one of the main limiting factors for large-scale deployment of ASIC supercomputers.
ASIC designs with higher flexibility are believed to have longer lifetimes and thus easier to amortize the NRE costs.
The main flexibility of Chiplet Cloud depends on the flexibility of chip design, which usually dominates in NRE.
It is feasible to redesign servers and software mapping for different generative language models using the same chip.
Since LLM scaling changes the number of parameters, while the operational intensity usually remains the same, Chiplet Cloud is able to support LLMs of larger sizes by adding chips.
LLMs may also have different element-wise operations, such as different activation functions and positional embeddings, our highly programmable SIMD cores are able to support all of these variations.

\begin{figure}[t]
    \centering
    \includegraphics[width=0.502\textwidth]{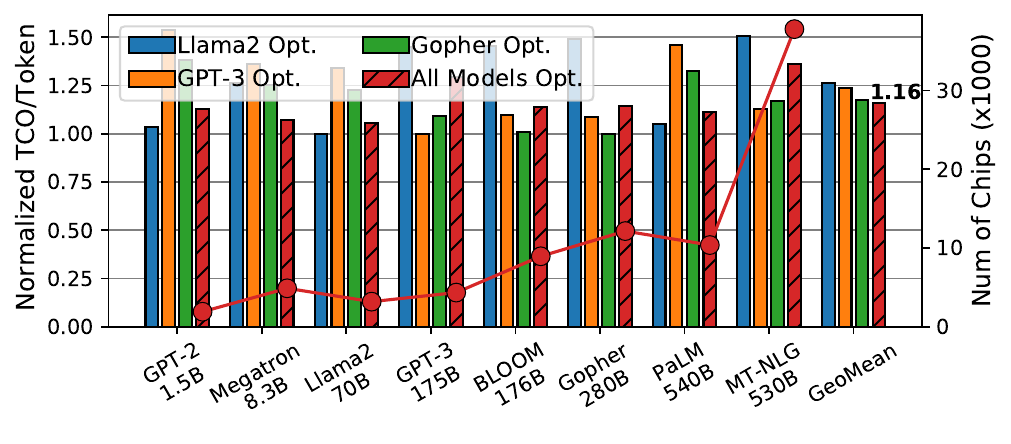}
    \caption{
     A Chiplet Cloud chip design is flexible to run models of different sizes via scale-up.
     Comparing to the model-optimized design, chip optimized for other models has TCO/TOken of $1.1\times$ to $1.5\times$ (blue, orange and green bar).
     One can also optimizes the chip for multi-model (dashed red bars) at only $1.16\times$ TCO/Token on average, and the number of chips used is shown in red dots.
    }
    \label{fig:flexibility}
\end{figure}

By adjusting the number of chips and optimizing the server and mapping, one chip design can to run models of different sizes without sacrificing too much TCO/Token.
In Figure \ref{fig:flexibility}, we show the impact on TCO/Token when mapping a chip to different models.
We first show 3 model-optimized chip designs in blue, orange and green bars for Llama2, Gopher, and GPT-3, respectively.
When running different models, it only increases TCO/Token by $1.1\times$ to $1.5\times$ compared to the corresponding model-optimized design.
When flexibility comes as the first priority, one can also set a multi-model optimization for the chip design. 
The red dashed box shows a design optimized for the geometric mean of TCO/Token on all 8 models, achieving an average overhead of only $0.16\times$ compared to the 8 single-model optimized designs.
The red dots represent the number of chips used for each model.
This demonstrates Chiplet Cloud has the flexibility to support various LLMs.

\subsection{NRE Discussion}
\label{sec:why_asic}

\begin{figure}
    \centering
    \includegraphics[width=0.42\textwidth]{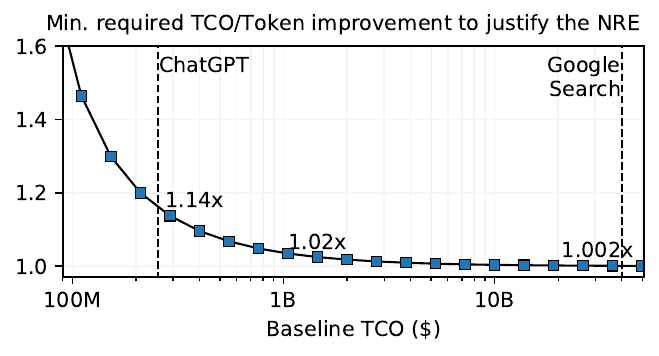}
    \caption{
    Designing ASIC for LLMs will be more cost-effective than using GPUs since the demand is so high.
    ChatGPT using GPUs has a TCO of \$255 M~\cite{chatgpt_cost_2023} per year, requiring only a $1.14\times$ TCO/Token improvement of ASIC to justify the NRE costs.
    }
    \label{fig:asic_profit}
\end{figure}

One major factor limiting the deployment of ASICs is non-recurring engineering (NRE) costs \cite{khazraee_moonwalk_2017}.
The barrier for overcoming NRE is primarily about the opportunity cost of running the workload on the current hardware offerings.
The difference in TCO between running a workload on an ASIC supercomputer vs the current available hardware platform determines the break even point for NRE, where the NRE cost directly equals the savings from using an ASIC supercomputer.
Figure~\ref{fig:asic_profit} shows the minimum required TCO/Token improvement in order to justify the NRE.
We extend the NRE model from Moonwalk~\cite{khazraee_moonwalk_2017} to use a 7nm technology node and estimate the NRE of an ASIC accelerator for large language models to be approximately \$35M, including silicon mask cost, CAD tools, IP licensing, flip-chip BGA packing, server designs, and labor.
Even if it were \$100M, the current cost of running workloads like ChatGPT and web search with integrated LLMs is so massive that it not only justifies the cost of creating ASIC supercomputers but going even further as to co-optimize those supercomputers for specifics LLMs for additional improvement in TCO per token.
This shows that the NRE cost of ASIC supercomputers is justifiable for modern workloads and thus customized hardware still remains the best solution to democratize LLMs.

\section{Related Work}

\textbf{Cloud-scale acceleration and TCO optimization.}
Microsoft's Brainwave \cite{fowers_configurable_2018} proposes FPGA-based clouds for ML.
Google's TPU \cite{jouppi_ten_2021} ML cloud is designed for performance per TCO.
Magaki et al \cite{magaki_asic_2016} proposes ASIC Clouds for scale-out application with TCO optimization.
Moonwalk \cite{khazraee_moonwalk_2017} optimizes the NRE for ASIC Clouds.

\textbf{Training and serving large language models.}
Megatron-LM~\cite{shoeybi_megatron-lm_2020} and DeepSpeed~\cite{aminabadi_deepspeed-_2022, rajbhandari_deepspeed-moe_2022} proposes multi-GPU training and inference solution to minimize latency while maximizing throughput.
\cite{narayanan_efficient_2021} improves pipeline parallelism and combines it with Megatron-LM and achieves high aggregate throughput.
PaLM~\cite{chowdhery_palm_2022} train a 540B parameter model on 6144 TPUv4 chips using Pathways~\cite{barham_pathways_2022}.
Pope et al \cite{pope_efficiently_2022} optimizes large scale transformer inference on TPUv4.

\textbf{ASIC accelerators for transformer models.}
ELSA~\cite{ham_elsa_2021} presents an approximation scheme for the attention mechanism.
SpAtten~\cite{wang_spatten_2021} exploits the token and head sparsity and quantization opportunities in the attention block.
EdgeBERT~\cite{tambe_edgebert_2021} leverages dynamic voltage-frequency scaling based on early exit prediction of ALBERT~\cite{lan_albert_2020}.
FLAT~\cite{kao_flat_2023} optimizes the dataflow in attention.
CTA~\cite{wang_cta_2023} proposes a HW-SW co-design for attention compression.


\section{Conclusion}
This paper presents Chiplet Cloud, a chiplet-based ASIC LLM-supercomputer architecture that achieves unprecedented TCO/Token for serving large generative language model.
It employs an on-chip memory architecture, CC-MEM, to eliminate bandwidth limitations, and a compression decoder for supporting sparse models, while moderating the die size to improve system costs.
We also propose a comprehensive design methodology that accurately explores Chiplet Cloud's design space.
We design Chiplet Cloud systems for eight language models 
and achieved up to $97\times$ and $18\times$ better TCO/Token compared to running on 
GPU and TPU cloud, respectively.
We believe Chiplet Cloud to be the best solution to democratise modern and future large generative language models.

\section*{Acknowledgements}
This work was supported by NSF Award 2118628.


\bibliographystyle{ACM-Reference-Format}
\bibliography{references}

\end{document}